# Design optimization of stochastic complex systems via iterative density estimation

Liu Wang-Sheng and Cheung Sai Hung

School of Civil and Environmental Engineering, Nanyang Technological University, Singapore

**Abstract**: Reliability-based design optimization (RBDO) provides a rational and sound framework for finding the optimal design while taking uncertainties into account. The main issue in implementing RBDO methods, particularly stochastic simulation based ones, is the computational burden arising from the evaluation of reliability constraints. In this contribution, we propose an efficient method which approximates the failure probability functions (FPF) to decouple reliability. Based on the augmentation concept, the approximation of FPF is equivalent to density estimation of failure design samples. Unlike traditional density estimation schemes, where the estimation is conducted in the entire design space, in the proposed method we iteratively partition the design space into several subspaces according to the distribution of failure design samples. Numerical results of an illustrative example indicate that the proposed method can improve the computational performance considerably.

## 1 Introduction

The primary task in engineering system design is to find the optimal solution by balancing the contradictory multidisciplinary requirements and the system performances. Reliability-based design optimization (RBDO) uses probability theory to explicitly take uncertainties associated with future excitations and system modelling into account, making it a unified and sound tool for aerospace and civil system design [1].

Roughly speaking RBDO is an optimization framework that incorporates reliability constraints. Here we denote a reliability constraint as $P_F(\boldsymbol{\varphi}) \leq [P_F]$ where $\boldsymbol{\varphi}$ is a vector of design variables, $P_F(\boldsymbol{\varphi})$ is the failure probability function (FPF) and $[P_F]$ is the allowable failure probability. That is, for a specific design configuration, the failure probability of its corresponding system should not exceed the allowable failure probability for a failure event $F$. Unfortunately, evaluation of reliability constraints is a daunting challenge due to the computational burden imposed by calculation of failure probability, particularly when complex numerical models are involved.

Over the last decade there has been considerable research efforts on proposing RBDO methods; here we describe two common ones. The first is an intuitive double-loop approach [2-4] that integrates reliability assessments into conventional deterministic optimization procedure. The second approach is to replace reliability constraints with deterministic constraints by approximating FPFs, thus decoupling reliability assessments from the optimization loop. Early efforts [5-8] restrict the class of FPFs to parametric functions, for instance, linear functions [6-8] or quadratic functions [5]. An obvious drawback of parametric functions is that if the underlying FPF cannot be not well modelled by a certain class of functions, then errors will be excessively large.

In this paper, we presented a novel method for approximating FPFs via density estimation. As pointed out by Au [9], approximation of FPFs can be described as a problem of density estimation. This paper is intended to show how the density function of failure samples can be accurately estimated by partitioning the design space, even for the small failure probability region.

## 2 Relation of the failure probability function and density function

Now consider an engineering system involving deterministic design variables $\boldsymbol{\varphi} = [\varphi_1, \cdots, \varphi_{n_\varphi}] \subset \mathbb{R}^{n_\varphi}$ and random variables $\boldsymbol{\Theta} = [\Theta_1, \cdots, \Theta_{n_\theta}] \subset \mathbb{R}^{n_\theta}$. In the augmented space $\Omega \subset \mathbb{R}^{n_\varphi + n_\theta}$ [9], design variables are artificially considered as random variables normally characterized by a uniform distribution $p(\boldsymbol{\varphi})$ supported on the ranges of $\boldsymbol{\varphi}$, then based on Bayes' Theorem the FPF can be expressed as

$$P_F(\boldsymbol{\varphi}) = P(F|\boldsymbol{\varphi}) = \frac{p(\boldsymbol{\varphi}|F)P(F)}{p(\boldsymbol{\varphi})} \propto p(\boldsymbol{\varphi}|F) \qquad (1)$$

As given in Equation (1), the FPF $P_F(\boldsymbol{\varphi})$ is proportional to the density function $p(\boldsymbol{\varphi}|F)$ as both $p(\boldsymbol{\varphi})$ and $P(F)$ are constants. This relationship is illustrated in Figure 1. Notice that $P(F)$ is rather the failure probability of the original system but the 'augmented' system after introducing artificial random variables.

To approximate the FPF, we first estimate $P(F)$ using direct Monte Carlo Simulation (dMCS) or Subset Simulation (SS). As a byproduct of this 'pilot simulation', a set of $N$ failure samples $\left\{\left[\boldsymbol{\varphi}_F^{(1)}, \boldsymbol{\Theta}_F^{(1)}\right], \cdots, \left[\boldsymbol{\varphi}_F^{(N)}, \boldsymbol{\Theta}_F^{(N)}\right]\right\}$ will be generated and its design component $\left\{\boldsymbol{\varphi}_F^{(1)}, \cdots, \boldsymbol{\varphi}_F^{(N)}\right\}$ are samples following $p(\boldsymbol{\varphi}|F)$ as shown in Figure 1. Then, the density function $p(\boldsymbol{\varphi}|F)$ is estimated based on simulated samples, which is the core part of the proposed method and will be discussed in detail in the next section. Finally, we scale $p(\boldsymbol{\varphi}|F)$ to obtain the FPF as suggested by Equation (1).

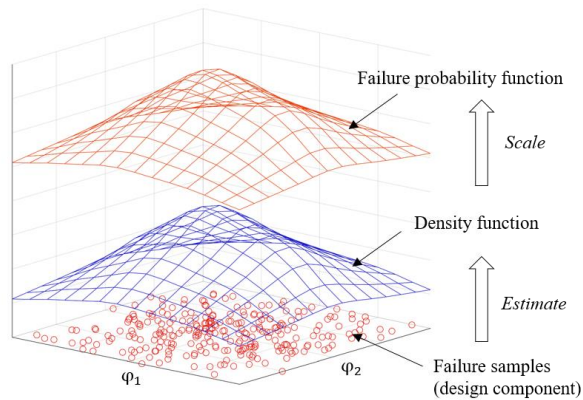

*Figure 1: Estimated density function and the corresponding failure probability function*

## 3 Iterative density estimation in partitioned design space

In this section, we will focus our attention on density estimation of $p(\boldsymbol{\varphi}|F)$ based on a finite number of samples. Histogram and kernel density estimation (KDE) are arguably the most prevalent techniques in engineering. But the application of histogram has been restricted to one-dimension case; KDE has long been criticized for its sensitivity to the choice of bandwidth and kernel types. Another issue is that as shown in Figure 1, most samples are scattering over the high density region rather than the low density region that we are interested in, resulting in large estimation errors.

We start with the detailed description of the proposed iterative scheme for density estimation in Section 3.1. In each iteration, failure samples are propagated from the high density region to low density region, and the density is estimated accordingly in the partitioned design space. We present how failure samples lying in a specific region are generated using Markov Chain Monte Carlo techniques in Section 3.2. We continue in Section 3.3 with Bayesian Sequential Partitioning (BSP) [10] to construct a piece-wise constant density estimator for each region, which will be smoothed by regression functions in Section 3.4.

## 3.1 An iterative scheme

The iterative scheme is illustrated in Figure 2 for a 2-d design space. In the pilot run, design component of failure samples $\left\{\boldsymbol{\varphi}_F^{(1)}, \cdots, \boldsymbol{\varphi}_F^{(N)}\right\}$ (for brevity, they are referred to as samples hereafter) are simulated over the entire design space $D_0$; then we use BSP to construct a piece-wise density estimator $\hat{p}_{D_0}(\boldsymbol{\varphi}|F)$ based on these samples. As the accuracy of the estimator is poor in the low density region, one should update the estimator in the following iterations. We define the low density region as $D_1 = \left\{\boldsymbol{\varphi}|\hat{p}_{D_0}(\boldsymbol{\varphi}|F) < p_0^*\right\} \subset D_0$ where $p_0^*$ is the partition density value for the pilot run. In this way, the initial design space is partitioned into two parts, i.e., high density region $S_1$ and low density region $D_1$. In iteration $k$, we start with simulating more samples in the low density region $D_k$. Notice from Figure 2 that additional samples can be sampled based on samples from previous simulation. We will discuss the sampler later in Section 3.2. Similarly, we partition the $D_k$ into $D_{k+1}$ and $S_{k+1}$ after obtaining $\hat{p}_{D_k}(\boldsymbol{\varphi}|F)$. The iteration process continues until the stopping criteria is satisfied. Ultimately, after $n_{it}$ iterations, the initial design space is partitioned into $n_{it} + 2$ mutually exclusive and collectively exhaustive regions $D_0 = S_1 \cup \cdots \cup S_{n_{it}+1} \cup D_{n_{it}+1}$.

By the Total Probability Theorem, the density function can be expressed as

$$p(\boldsymbol{\varphi}|F) = p_{D_0}(\boldsymbol{\varphi}|F) = \sum_{k=1}^{n_{it}+1} p_{S_k}(\boldsymbol{\varphi}|F)P(S_k|F) + p_{D_{n_{it}+1}}(\boldsymbol{\varphi}|F)P(D_{n_{it}+1}|F)$$
$$= \sum_{k=0}^{n_{it}} p_{D_k}(\boldsymbol{\varphi}|F)P(D_k|F) \quad (2)$$

Recall that in iteration $k$, $D_{k+1}$ is defined as the low density region. So the ratio of probability of samples lying in $D_{k+1}$ and $D_k$ is given by

$$\frac{P(D_{k+1}|F)}{P(D_k|F)} = P\left(p_{D_k}(\boldsymbol{\varphi}|F) < p_k^*\right) = P_k^*, \quad k = 0, \cdots, n_{it} \quad (3)$$

Then,

$$P(D_k|F) = \prod_{j=0}^{k-1} P_j^*, \quad k = 1, \cdots, n_{it} \quad (4)$$

Either the value of the partition density value $p_k^*$ or the probability ratio $P_k^*$ can be assigned by users.

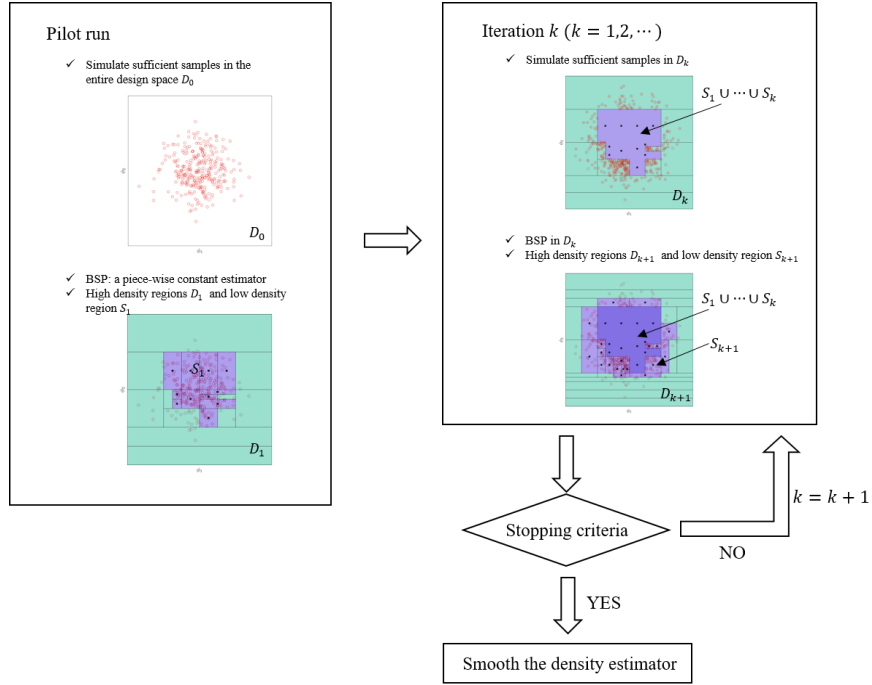

*Figure 2: An iterative scheme for density estimation*

## 3.2 The sampler for failure sample generation

At the beginning of each iteration, additional samples following $p_{D_k}(\boldsymbol{\varphi}|F)$ should be simulated for density estimation over $D_k$. We use modified Metropolis-Hastings (MMH) algorithm, a Markov Chain Monte Carlo (MCMC) [11] algorithm, for obtaining a sequence of samples from our target distribution $p_{D_k}(\boldsymbol{\varphi}|F)$.

## 3.3 Bayesian Sequential Partitioning

Unlike regular partitioning in conventional histogram, in BSP the resulting density estimators are supported by binary partitions of the space of interest.

For brevity, we denote the space of interest $D_k$ as $D$, samples $\{\boldsymbol{\varphi}_F^{(1)}, \cdots, \boldsymbol{\varphi}_F^{(N)}\}$ as $\{\boldsymbol{\varphi}\}$ and the estimated density function $\hat{p}_{D_k}(\boldsymbol{\varphi}|F)$ as $\hat{p}(\boldsymbol{\varphi})$. Let a partition $x_t$ be a set of disjoint subspaces $A_i$ of $D$ where $D = \cup_{i=1}^t A_i$. The piece-wise constant estimator $\hat{p}(\boldsymbol{\varphi})$ will be $\hat{p}(\boldsymbol{\varphi}) = \cup_{i=1}^t p_i I_{A_i}(\boldsymbol{\varphi})$ where $I_{A_i}(\boldsymbol{\varphi})$ is an indicator function which equals to 1 when $\boldsymbol{\varphi} \in A_i$ and 0 otherwise. And $p_i = \theta_i/|A_i|$ where $|A_i|$ is the volume of $A_i$, $\theta_i$ represents the probability of $\boldsymbol{\varphi} \in A_i$.

Assume that the prior of a partition $x_t$ is proportional to $e^{-\beta t}$, its posterior distribution given samples $\{\boldsymbol{\varphi}\}$ is given by [10]

$$P(x_t|\{\boldsymbol{\varphi}\}) \propto e^{-\beta t} \frac{B(n_1 + \alpha, \cdots, n_t + \alpha)}{B(\alpha, \cdots, \alpha)} \prod_{i=1}^t \frac{1}{|A_i|^{n_i}} \qquad (5)$$

where $\alpha$ is a constant parameter for a Dirichlet distribution, $B(\cdot)$ is a Beta function, and $n_i$ is the number of samples in $A_i$. The partitions $x_t$ are simulated using Sequential Importance Sampling [12] through generating a binary cut at each level as illustrated in Figure 3.

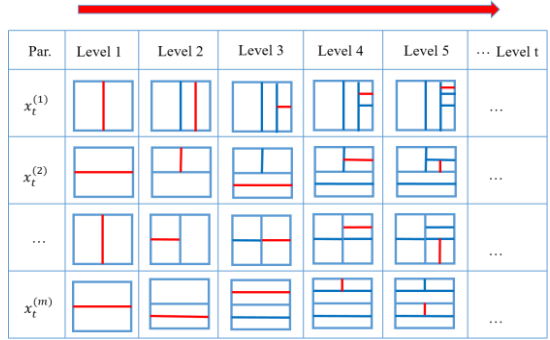

*Figure 3: Sequential partitioning process of BSP (in each level, a total number of m partitions are simulated)*

Partitions by BSP are data-adaptive, making its computation complexity linear to the sample size and sample dimension. Additionally, unlike to kernel density estimation, which suffers from the well-known 'boundary bias' problem, BSP will not introduce the boundary bias even when complex boundaries are involved.

### 3.4 Smoothing using regression functions

As BSP only gives piece-wise constant estimators, the density values for adjacent subspaces are discontinuous. But in some applications, such as design sensitivity analysis, the property of continuity is preferred.

To smooth the piece-wise constant estimator, we combine the support points obtained from the high density regions with appropriate regression functions. As shown in Figure 4, centers of each rectangle in the high density region are considered as support points and their corresponding density values are calculated using BSP described in Section 3.3.

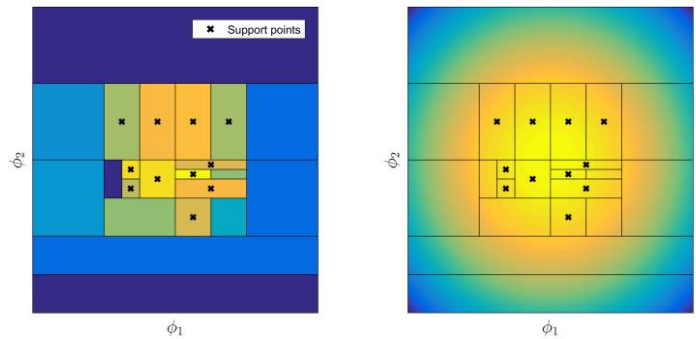

*Figure 4: Smoothing of a piece-wise constant estimator (left panel: piece-wise constant estimator; right panel: smoothed estimator)*

A wide range of parametric or nonparametric regression functions [13] can be chosen, including Gaussian processes, neural networks, moving least squares, etc. Choosing a regression function for a particular problem requires the comparison of functions within a particular family and across different function families, typically referred to as a model selection problem. Model selection is beyond the scope of this paper.

# 4 An illustrative example

In this section, we apply the proposed DOPADS to an artificial RBDO problem of a cantilever beam. This example is intended to illustrate strengths of the proposed method in approximating nonlinear FPFs.

## 4.1 Description

Consider a cantilever box beam shown in Figure 5. The length of the beam is $500\ mm$. Model parameters are taken as random variables, i.e., $\boldsymbol{\theta} = [b, h, t, \rho, E]$ where $b$, $h$, and $t$ are the width, height and thickness of the cross section respectively, $\rho$ is the density and $E$ is Young's modulus. We take the mean values of width $\bar{b}$ and the height $\bar{h}$ as design variables, i.e., $\boldsymbol{\varphi} = [\bar{b}, \bar{h}]$. The intervals of design variables and statistic characteristics of random variables are summarized in Table 1.

Our objective is to minimize the mean cross-sectional area of the beam. In the reliability constraint, the failure event $F$ is defined as the first natural frequency $\omega_1$ falling into the frequency interval that should be avoided, i.e., $550 \leq \omega_1(\boldsymbol{\varphi}, \boldsymbol{\theta}) \leq 600\ (rad/s)$.

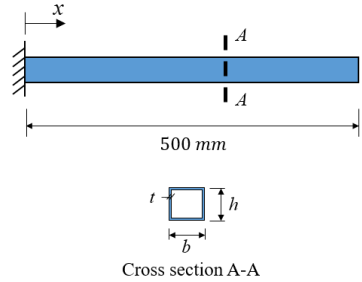

*Figure 5: Diagram of the studied cantilever box beam*

*Table 1: Design variables and random variables*

|  | Parameter | Interval | Distribution | Mean | S.d. |
|---|---|---|---|---|---|
| Design variables | Mean of width $\bar{b}$ $(mm)$ | [30,50] | -- | -- | -- |
|  | Mean of height $\bar{h}$ $(mm)$ | [30,50] | -- | -- | -- |
| Random variables | Width $b$ $(mm)$ | -- | Normal | $\bar{b}$ | $0.02 \times \bar{b}$ |
|  | Height $h$ $(mm)$ | -- | Normal | $\bar{h}$ | $0.02 \times \bar{h}$ |
|  | Thickness $t$ $(mm)$ | -- | Normal | 2.0 | 0.1 |
|  | Young's modulus $E$ $(GPa)$ | -- | Normal | 210 | 4.2 |
|  | Density $\rho$ $(kg/m^3)$ | -- | Normal | 7800 | 156 |

## 4.2 Approximated failure probability function

After the pilot run and three iterations, the stopping criteron was satisfied as the corresponding failure probability of partition density value $p_3^*$ in the third iteration is smaller than the failure probability we are interested in, i.e., $10^{-4}$. As shown in Figure 6, the initial design space was ultimately partitioned into five subspaces from high density regions to low density regions. In each subspace, density values of support points were calculated.

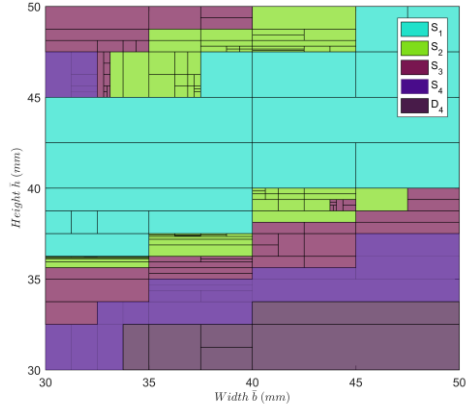

*Figure 6: Partitioned design space after three iterations*

We used Gaussion processes as a regression function to fit the density function and scaled it to obtain the FPF based on Equation (1). The result was compared with that using direct Monte Carlo Simulation (dMCS). That is, we uniformly meshed the design spaced into a grid of 21 by 21, then calculated the failure probability for each gridpoint. As shown in Figure 7, FPFs using these two methods are consistent with each other, indicating a good accuracy of the proposed method. The approximated function provides insights to various RBDO problems including sensitivity analysis and feasible region identification. For instance, the gradient of the approximated function can be plotted as shown in Figure 8.

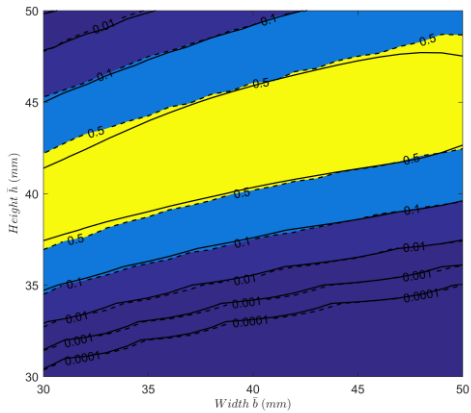

*Figure 7: Contour plot of failure probability functions*

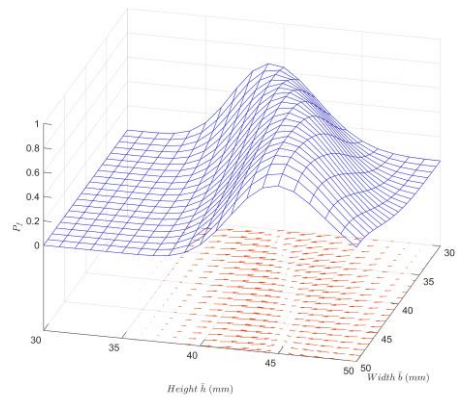

*Figure 8: Approximated failure probability funciton and the vector field of its gradient*

## 4.3 Optimal solutions

Deterministic nonlinear optimization was used after approximating FPFs. Optimal solutions are summarized in Table 2 for three cases with different allowable failure probabilities $[P_F] = 0.01, 10^{-3}, 10^{-4}$. The efficiency of the proposed method can be measured by the total number of model performance evaluations. In this example, it requires a total number of 36000 evaluations in the proposed method and $5.68 \times 10^7$ for dMCS. The proposed method is efficient compared to exhaustively performing dMCS for each gridpoint while optimal solutions stay close as shown in Table 2. Note that with FPFs at hand, we are not bothered to redo the reliability analysis when the allowable failure probabilities are changed by decision makers.

Table 2: Optimal solutions for diffrent allowable failure proabilities

| $[P_F]$ | The proposed method $[\bar{b}, \bar{h}](mm)$ | dMCS $[\bar{b}, \bar{h}](mm)$ |
|---|---|---|
| 0.01 | [30.0, 32.9] | [30.0, 32.7] |
| $10^{-3}$ | [30.0, 31.5] | [30.0, 31.4] |
| $10^{-4}$ | [30.0, 30.4] | [30.0, 30.4] |

# 5  Conclusions and future works

In this contribution, we presented an efficient decoupling method for RBDO problems which approximates the failure probability function based on iterative density estimation. We gain the efficiency mainly from the avoidance of repeated reliability assessments by generating failure samples from low density regions to high ones and estimating their density distributions.

In the future studies, we will conduct computer experiments for cases with more than two design variables.

# References

[1] Tsompanakis Y, Lagaros ND, Papadrakakis M. Structural Design Optimization Considering Uncertainties: Structures & Infrastructures Book, Vol. 1, Series, Series Editor: Dan M. Frangopol: CRC Press; 2008.
[2] Tu J, Choi KK, Park YH. A new study on reliability-based design optimization. Journal of Mechanical Design. 1999;121:557-64.
[3] Royset J, Polak E. Reliability-based optimal design using sample average approximations. Probabilistic Engineering Mechanics. 2004;19:331-43.
[4] Papadrakakis M, Lagaros ND. Reliability-based structural optimization using neural networks and Monte Carlo simulation. Computer Methods in Applied Mechanics and Engineering. 2002;191:3491-507.
[5] Gasser M, Schuëller GI. Reliability-based optimization of structural systems. Mathematical Methods of Operations Research. 1997;46:287-307.
[6] Jensen HA. Structural optimization of linear dynamical systems under stochastic excitation: A moving reliability database approach. Computer Methods in Applied Mechanics and Engineering. 2005;194:1757-78.
[7] Ching J, Hsieh YH. Approximate reliability-based optimization using a three-step approach based on subset simulation. Journal of Engineering Mechanics. 2007;133:481-93.
[8] Ching J, Hsieh YH. Local estimation of failure probability function and its confidence interval with maximum entropy principle. Probabilistic Engineering Mechanics. 2007;22:39-49.
[9] Au SK. Reliability-based design sensitivity by efficient simulation. Computers and Structures. 2005;83:1048-61.
[10] Lu L, Jiang H, Wong WH. Multivariate density estimation by bayesian sequential partitioning. Journal of the American Statistical Association. 2013;108:1402-10.
[11] Au S-K, Beck JL. Estimation of small failure probabilities in high dimensions by subset simulation. Probabilistic Engineering Mechanics. 2001;16:263-77.
[12] Liu JS. Monte Carlo strategies in scientific computing: Springer Science & Business Media; 2008.
[13] Kleinbaum D, Kupper L, Nizam A, Rosenberg E. Applied regression analysis and other multivariable methods: Cengage Learning; 2013.